\documentclass[a4paper,fleqn]{cas-dc}

\newcommand{\mnras}{MNRAS}
\newcommand{\apj}{ApJ}
\newcommand{\apjl}{ApJL}
\newcommand{\apjs}{ApJS}
\newcommand{\aj}{AJ}
\newcommand{\aap}{A\&A}
\newcommand{\araa}{ARA\&A}

\newcommand{\nat}{Nature}
\usepackage[authoryear]{natbib}

\def\tsc#1{\csdef{#1}{\textsc{\lowercase{#1}}\xspace}}
\tsc{WGM}
\tsc{QE}
\begin{document}
\let\WriteBookmarks\relax
\def\floatpagepagefraction{1}
\def\textpagefraction{.001}

\shorttitle{From individual element abundances to star formation histories}    

\shortauthors{M. Longhetti et al.}  

\title [mode = title]{Quiescent galaxies at cosmic noon: from individual element abundances to star formation histories}  

\tnotemark[1] 

\tnotetext[1]{This paper is based on a contribution presented at the meeting “Unveiling the Universe with SHARP”, held in Milan from 30 September to 2 October 2024} 


\author[1]{M. Longhetti}[orcid=0000-0002-6142-4822]%[<options>]

\cormark[1]


\ead{marcella.longhetti@inaf.it}


\credit{Conceptualization, Methodology, Writing -  original draft, Writing - review \& editing}


\affiliation[1]{organization={INAF - Osservatorio Astronomico di Brera},
            addressline={via Brera, 28}, 
            city={Milano},
%          citysep={}, % Uncomment if no comma needed between city and postcode
            postcode={20121}, 
            %state={},
            country={Italy}}
            
\affiliation[2]{organization={INAF - Osservatorio Astronomico di Roma},
            addressline={via Frascati, 33}, 
            city={Monte Porzio Catone},
%          citysep={}, % Uncomment if no comma needed between city and postcode
            postcode={00078}, 
            %state={},
            country={Italy}}

\affiliation[3]{organization={INAF - Astronomical Observatory of Trieste},
            addressline={via Beirut, 2}, 
            city={Trieste},
%          citysep={}, % Uncomment if no comma needed between city and postcode
            postcode={34151}, 
            %state={},
            country={Italy}}

\affiliation[4]{organization={IFPU - Institute for Fundamental Physics of the Universe},
            addressline={via G.B. Tiepolo, 11}, 
            city={Trieste},
%          citysep={}, % Uncomment if no comma needed between city and postcode
            postcode={34143}, 
            %state={},
            country={Italy}}

\affiliation[5]{organization={INAF-Osservatorio Astrofisico di Arcetri},
            addressline={Largo Enrico Fermi, 5}, 
            city={Firenze},
%          citysep={}, % Uncomment if no comma needed between city and postcode
            postcode={50126}, 
            %state={},
            country={Italy}}

\affiliation[6]{organization={INAF – IASF Milano},
            addressline={via A. Corti, 12}, 
            city={Milano},
%          citysep={}, % Uncomment if no comma needed between city and postcode
            postcode={20133}, 
            %state={},
            country={Italy}}

\affiliation[7]{organization={INAF - Osservatorio Astronomico di Capodimonte},
            addressline={ Via Moiariello, 16}, 
            city={Napoli},
%          citysep={}, % Uncomment if no comma needed between city and postcode
            postcode={80131}, 
            %state={},
            country={Italy}}

\affiliation[8]{organization={Institute of Cosmology, University of Portsmouth},
            addressline={Burnaby Road}, 
            city={Portsmouth},
%          citysep={}, % Uncomment if no comma needed between city and postcode
            postcode={PO1 3FX}, 
            %state={},
            country={UK}}

\affiliation[9]{organization={School of Mathematics and Physics, University of Portsmouth},
            addressline={Lion Gate Building}, 
            city={Portsmouth},
%          citysep={}, % Uncomment if no comma needed between city and postcode
            postcode={PO1 3HF}, 
            %state={},
            country={UK}}

\affiliation[10]{organization={School of Physics and Astronomy, University of Leeds},
            addressline={Bragg Building},
            city={Leeds},
%          citysep={}, % Uncomment if no comma needed between city and postcode
            postcode={LS2 9JT}, 
            %state={},
            country={UK}}

\author[1]{F. R. Ditrani}%[]
\credit{Conceptualization, Methodology, Writing -  original draft, Writing - review \& editing}
\author[2]{D. Bevacqua}
\credit{Writing - review \& editing}
\author[3,4]{G. De Lucia}
\credit{Writing review \& editing}
\author[5]{A. Gallazzi}
\credit{Writing - review \& editing}
\author[6]{A. Gargiulo}
\credit{Writing - review \& editing}
\author[7]{F. La Barbera}
\credit{Writing - review \& editing}
\author[8]{C. Maraston}
\credit{Writing - review \& editing}
\author[9,10]{D. Thomas}
\credit{Writing - review \& editing}
\author[6]{P. Franzetti}
\credit{Software, Writing - review \& editing}
\author[6]{C. Mancini}
\credit{Writing - review \& editing}
\author[5]{L. Scholz-Díaz}
\credit{Writing - review \& editing}
\author[5]{S. Zibetti}
\credit{Writing - review \& editing}
% Footnote of the second author

% Email id of the second author
%\ead{f.ditrani1@campus.unimib.it}


%\affiliation[2]{organization={},
%          citysep={}, % Uncomment if no comma needed between city and postcode

\cortext[1]{Corresponding author}



\begin{abstract}
Understanding the formation and early evolution of massive quiescent galaxies is crucial for a complete picture of galaxy formation in a cosmological context. Indeed, in a cold dark matter dominated Universe hierarchical galaxy formation makes it difficult to form massive galaxies quickly and preserve them in just a passive evolution state for most of the cosmic time, as suggested by studying their local fossil population. 
Such a comprehensive picture can be obtained through precise measurements of individual elemental abundances, which encode the timescales and physical conditions of their star-formation and assembly histories. While extensive work at low and intermediate redshift has revealed strong correlations between stellar mass, metallicity, and abundance ratios, studies at $z \sim 2-3$ (i.e., the epoch when most progenitors of present day early-type first quenched) remain extremely limited. This is primarily due to the faintness and compactness of these systems, which make it extremely challenging to obtain medium-resolution near-infrared spectra with the quality required for a detailed chemical analysis. We present a science case to measure full elemental abundance patterns in a representative sample of massive quiescent galaxies at cosmic noon, exploiting rest-frame optical absorption features.
We show that deriving robust stellar population constraints for these systems requires spectra with S/N $\gtrsim 20$ per resolution element, together with a spatial resolution comparable to their $\sim 0.1$ arcsec sizes. We show, through dedicated exposure time simulations, that such requirements can only be met with a 40m class telescope equipped with a powerful adaptive optics system. In this context, SHARP-NEXUS, in combination with MORFEO-MCAO capabilities, emerges as the only viable ELT instrument capable of enabling this science, providing both the sensitivity and angular resolution needed. 

%Although the low target density limits multiplexing efficiency, this can be mitigated by observing galaxy overdensities at $z>2$ or by coordinating observations with complementary programmes targeting star forming galaxies at similar redshifts. Recent examples from high-redshift clusters and protoclusters, as well as the operational model of multi survey multiplex sharing in facilities such as 4MOST, demonstrate the feasibility and scientific potential of such strategies. SHARP-NEXUS would therefore open an entirely new observational window onto the chemical enrichment histories of the earliest massive galaxies.

Although massive quiescent galaxies are rare at these redshifts, their study can efficiently exploit the instrument’s multiplexing by targeting galaxy overdensities and coordinating with surveys of star-forming galaxies. Such synergistic strategies, proven feasible in high-redshift clusters and by multiplex-sharing facilities like 4MOST, would allow SHARP-NEXUS to open a new window on the chemical enrichment of the earliest massive galaxies.

%\nocite{*}%% Remove this line from your manuscript.
\end{abstract}


%\item 
%\item 
%\item 

%\nocite{*}

% Keywords

\maketitle

% Main text

% Use the style of numbering in square brackets.
% If nothing is used, default style will be taken.
%\begin{enumerate}[a)]
%\item 
%\item 
%\item 
%\end{enumerate}  

%\begin{itemize}
%\item 
%\item 
%\item 
%\end{itemize}  

%\begin{description}
%\item[]
%\item[] 
%\item[] 
%\end{description}  

%\clearpage %%Remove this from your manuscript

% Uncomment and use as the case may be
%\begin{theorem} 
%\end{theorem}

% Uncomment and use as the case may be
%\begin{lemma} 
%\end{lemma}


\section{Introduction}\label{intro}
Understanding how massive quiescent galaxies form and evolve is key to a complete picture of galaxy formation in a cosmological context. In a Universe dominated by cold dark matter, hierarchical assembly makes it challenging to build these galaxies rapidly and maintain them in a predominantly passive state over most of cosmic time, as revealed by the fossil record of local systems (e.g. \citealt{carnall2023}; \citealt{valentino2023}). Their well-behaved redshift evolution also makes them valuable cosmological probes (e.g., BOSS, \citealt{eisenstein2005}). By measuring the elemental abundances of their stellar populations, we can probe the physical processes that governed their assembly.
Indeed, the chemical composition of a galaxy results from the intricate balance between several physical processes, including the accretion of pristine gas, the expulsion of metal-enriched material through galactic winds, the efficiency of star formation, the cumulative effect of mergers, and ultimately, the stellar initial mass function  (IMF). While the metallicity of the interstellar medium traces the chemical state of the galaxy at the epoch probed by the observations, the elemental abundances locked in its stars record the integrated history of chemical enrichment throughout its formation and evolution (e.g., \citealt{thomas2010}). For this reason, stellar abundance patterns provide a powerful diagnostic of both the star formation and assembly pathways of galaxies (e.g., \citealt{dere} and references therein). In particular, for quiescent systems, where strong stellar absorption lines dominate the spectra, stellar chemical compositions serve as especially sensitive tracers of their past star forming activity.

In the early Universe, only hydrogen, helium, and trace amounts of lithium were produced through primordial nucleosynthesis. The formation of heavier elements, collectively referred to as \textit{metals}, began only with the first generation of stars. Inside them, thermonuclear reactions progressively synthesise heavier elements, with the yield depending on stellar mass and evolutionary stage. Massive stars ($M \ge 8M_\odot$) evolve rapidly ($\sim 1-30$ Myr) and end their lives as core-collapse supernovae (Type II SNe), enriching the interstellar medium (ISM) with so-called $\alpha$-elements (e.g., O, Mg, Si, Ca, Ti). In contrast, Type Ia supernovae, which originate from long-lived, lower-mass progenitors in binary systems, are the main producers of iron peak elements, and release their products on longer timescales (e.g., $\sim 1$ Gyr\  \citealt{greggio2005}; \citealt{nomoto}). This delayed enrichment plays a crucial role, as the ratio between $\alpha$-elements and iron traces the timescale over which star formation occurred in a galaxy\ (\citealt{thomas2005epochs}).

Analyses of stellar populations and chemical abundances in galaxies at $z \sim 0$ have provided key insights into the formation pathways and evolutionary histories of massive quiescent systems. A well established result is that the most massive early-type galaxies in the local Universe are also the most metal enriched (e.g., \citealt{trager2000stellar}; \citealt{gallazzi2005ages}; \citealt{thomas2010}). This correlation is generally interpreted as a consequence of the galaxy gravitational potentials: in deeper potential wells, stellar winds and supernova-driven outflows are less efficient at expelling the metal enriched gas (e.g., \citealt{tremonti2004} and references therein).

Massive quiescent galaxies are also observed to host the oldest stellar populations and to exhibit enhanced $\alpha$-element abundances, commonly quantified through the [$\alpha$/Fe] ratio (e.g., \citealt{thomas2005epochs}; \citealt{mcdermid2015}; \citealt{conroy2018metal}). These findings point toward a scenario in which the most massive galaxies assembled their stellar content early and over very short timescales (\citealt{delarosa11}).
Moreover, detailed studies of individual elemental abundance patterns have shed light on the nucleosynthetic origin of various elements in massive systems. For instance, the Fe peak element cobalt appears to receive a non negligible contribution from core collapse supernovae (\citealt{conroy2013modeling}), whereas $\alpha$-elements such as calcium and titanium show evidence of significant production in Type Ia events (e.g., \citealt{saglia2002}; \citealt{Thomas2003}; \citealt{Johansson2012}; \citealt{labarbera2025}).

Surveys targeting galaxies at redshifts $0.5 < z < 1.2$, (e.g., LEGA-C, \citealt{van2016vlt}; VANDELS, \citealt{carnall2019}). have provided an important bridge between the early universe and local quiescent systems. Yet, even at these epochs, roughly halfway through cosmic history, the interpretation remains complex due to processes such as rejuvenation episodes and morphological transformations. Establishing a direct evolutionary connection with present-day early-type is further complicated by the so called progenitor bias. 
Since galaxies quench at different cosmic times, the passive population continuously grows, meaning that passive galaxies observed at high redshift correspond only to the earliest quenched systems. Therefore, they are not necessarily representative of all passive galaxies observed in the local Universe.(\citealt{vandokkum2001}). 
Some galaxies cease star formation earlier and stay compact, while others quench later, typically exhibiting larger sizes. Consequently, to best constrain the mechanisms that shaped quiescent galaxies, it is essential to observe them near the time of their formation and quenching, around redshift $z \sim 2$, before subsequent mergers alter their stellar populations (\citealt{Barro2013}, \citealt{Sampaio25}). 
At that epoch a substantial fraction of the progenitors of the present-day passive galaxy population is thought to have already undergone quenching, making this epoch particularly suitable for investigating the stellar populations of the progenitors of massive quiescent galaxies in the local Universe. Indeed, the extent to which progenitor bias affects the interpretation of the results obtained for z $\sim$ 2 samples remains debated. Importantly, systematic differences in the stellar population models and analysis techniques adopted at different redshifts can themselves introduce significant variations in the inferred stellar ages. For example, \cite{ditrani26} showed that systematic differences of up to 1.5-2 Gyr in the inferred ages of massive quiescent galaxies at z < 0.8 can emerge simply from the choice of different evolutionary models adopted for the spectral analysis, even when the same galaxies, datasets, and spectral indicators are considered. By performing a homogeneous analysis of BOSS and DESI data for galaxies spanning z $\sim$ 0.8 to the local Universe, they found that the stellar populations of intermediate redshift massive quiescent galaxies are consistent with quenching epochs at z$>$1.5-2, depending on the adopted spectrophotometric model. These results suggest that, while progenitor bias cannot be completely neglected, its impact on the interpretation of the massive quiescent population at z$\sim$ 2 is likely negligible.

Until recently, observational limitations, particularly in the pre-JWST era, have made such studies virtually impossible. From the ground, the main restriction has been telescope size, because even with 10-meter class facilities, obtaining spectra with sufficient signal to noise ratios (S/N) for stellar population studies (i.e., $\gtrsim 20/$\AA, \citealt{zibetti24}) is unfeasible for galaxies at $z \sim 2$, whose typical magnitudes are around K$_{AB} \sim 24$ (see next section). The advent of JWST has partially mitigated this issue. Although its mirror is \textit{only} $6.5$ meters in diameter, its space based operation provides a crucial advantage, that is the absence of atmospheric background, which strongly limits near infrared observations from the ground.

However, JWST observations are primarily limited by the quality of the available spectra for detailed chemical abundance measurements. Indeed, only very few dedicated surveys (e.g., SUSPENSE, \citealt{slob2024}; DeepDive, \citealt{ito2025}) have obtained spectra suitable for this purpose, targeting only a handful of galaxies and mostly at $z < 2.5$.  A comprehensive study of the formation of quiescent galaxies requires samples encompassing a variety of environments, stellar masses, and structural properties, ideally including at least several dozen galaxies and extending to higher redshift.

The paper is organised as follows. In Sect.~\ref{proposal}, we present the observational proposal, describing the target sample, the observational requirements, and the theoretical models relevant to achieving its scientific goals. In Sect.~\ref{sharp}, we demonstrate the suitability of SHARP-NEXUS for this programme, while Sect.~\ref{obsstra} details the proposed observing strategy. Finally, in Sect.~\ref{summary}, we summarise the main aspects of the proposed science case.

Throughout the paper, we adopt a standard $\Lambda$CDM cosmology with $\Omega_M = 0.3089$, $\Omega_\Lambda = 0.6911$, and $H_0 = 67.8$ km s$^{-1}$ Mpc$^{-1}$.  Magnitudes are in the AB system \citep{oke1974absolute}. 
%However, JWST observations are inherently limited in sample size. The number of galaxies that can be observed remains small, preventing systematic and statistically robust analyses. A comprehensive study of the formation of quiescent galaxies requires samples encompassing a variety of environments, stellar masses, and structural properties, ideally including at least several dozen galaxies.

\section{Proposed study}\label{proposal}
We propose to perform a systematic study of a mass complete sample of massive quiescent galaxy population at cosmic noon, aiming to derive the abundances of individual chemical elements in their stellar component. Our selection is specifically designed to be representative of the z $\sim$ 2 massive passive galaxy population, which constitutes the main progenitor population of present-day massive quiescent galaxies. The analysis will be based on the comparison of selected continuum absorption features with predictions from stellar population models as a function of chemical composition. Similar studies on samples of galaxies in the local Universe (\citealt{Johansson2012}; \citealt{conroy2018metal}; \citealt{feldmeierlonoce}) and at intermediate redshift (\citealt{beverage2023}; \citealt{bevacqua2023}) have proven successful in recovering detailed stellar chemical abundances. The aim of this project is to extend these investigations to higher-redshift galaxies, close to the epochs of their formation and quenching, before subsequent evolution dilutes the signatures imprinted in their original stellar populations.

\subsection{The sampled galaxies}\label{sample}
The target sample will consist of massive quiescent galaxies at $z \simeq 2-3$. According to \cite{schreiber2018}, galaxies with $\log (M_\star/M_\odot) > 10.5$ are expected to have apparent $K_{A}$ magnitudes brighter than $24$. Moreover, based on \cite{ito2024}, such galaxies typically exhibit effective radii of about $0.1$ arcseconds, or even smaller. 

To identify the quiescent population, candidate targets will be pre-selected based on their rest-frame colours, using the U-V versus V-J colour-colour diagram (UVJ selection; e.g. \citealt{williams2009}; \citealt{Straatman2016}). We will also consider the $(ugi)_{s}$ selection proposed by \cite{Antwi-Danso2023}, which is designed to reduce contamination from dust-obscured star-forming galaxies and systems with strong emission lines.

In Section \ \ref{obsstra}, we describe in more detail the observing strategy of the survey we propose to carry out with the multi-object spectrograph SHARP–NEXUS (\citealt{saracco26sharp}, see also https://sharp.brera.inaf.it/instrument-design/).

\subsection{The required wavelength range}\label{wrange}
The rest-frame optical range is the key spectral window for our purposes. Experience from past surveys teaches us that the interval between 4000\ \AA\  and 6000\ \AA\  is particularly valuable, as it includes absorption lines due to several chemical elements (e.g., Fe, Mg, Ca, C - see Figure 1), originating from the stars that dominate the stellar mass of galaxies. (e.g., \citealt{Thomas2003}; \citealt{gallazzi2005ages} \citealt{maraston2011}; \citealt{lonoce2020}). 
%They are well understood and have been studied over decades. Stellar population models are well calibrated with globular clusters (Maraston et al 2003; Thomas et al 2003; Thomas et al 2011; etc) and local galaxies.

Although UV absorption features can also provide useful information on chemical abundances, they are primarily sensitive to the younger stellar components and to the possible presence of hot evolved stars (e.g. P-AGB and blue HB stars, see \citealt{le2016modelling}), and they can be strongly affected by the presence of even a tiny fraction (i.e. $<0.1\%$  of the overall mass) of very young (i.e. age $<100$ Myr) and hot stars, with drastic changes seen in the continuum and spectral features (e.g., \citealt{salvador2021young}).  Since these components are expected  to dominate the UV spectrum, UV diagnostics are less suitable for tracing the bulk stellar populations in massive quiescent galaxies.

From the literature, it also emerges that a spectral resolution of R $\gtrsim 2000$ and a S/N ratio $\gtrsim 20$ per resolution element are generally required to robustly constrain the stellar population properties of galaxies, such as ages and metallicities (e.g., \citealt{gallazzi2005ages}; \citealt{ditrani2025}). 
With spectra of the quality described above, it is possible to obtain constraints with upper-limit uncertainties of about $0.1$ dex in metallicity and $0.5$ Gyr in age. This is further facilitated by the fact that, at the redshifts considered, the Balmer absorption lines are particularly deep and highly sensitive to the age of the stellar population. Indeed, the age of the Universe at such high redshifts implies that the stellar populations cannot be older than $\sim 2$ Gyr, allowing us to exploit the regime in which the Balmer indices are most sensitive to age (see for example Figure 1 - left panel - in \citealt{ditrani23steps}).

% Figure
\begin{figure*}\label{indici}%[]
  \centering
   \includegraphics[width=15cm]{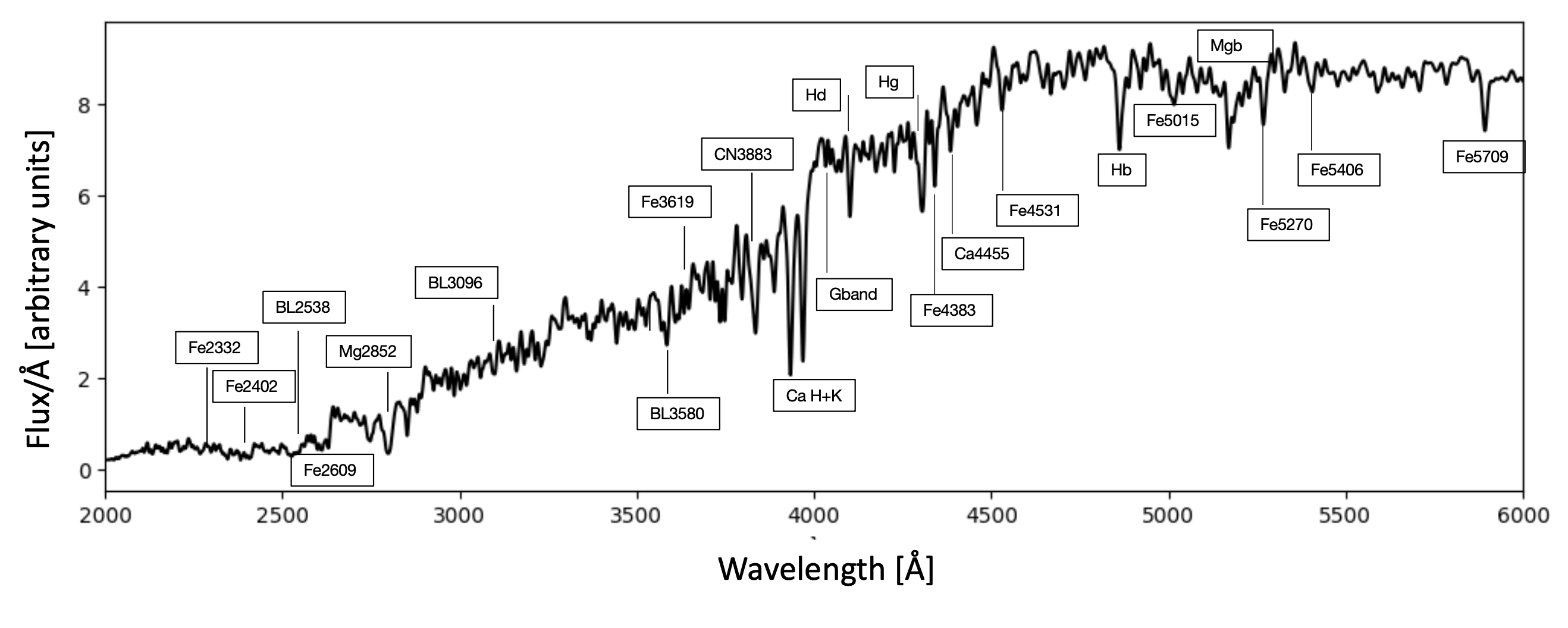} 
    \caption{Synthetic stellar continuum of a 2 Gyr old stellar population with solar metallicity. As can be seen, the optical portion of the spectrum contains numerous absorption lines that are useful for deriving the stellar population properties of galaxies.}
\end{figure*}

\subsection{Theoretical models}\label{models}

In order to convert absorption line strengths (measured as spectral indices) into element abundances, we need suitable models that predict the expected spectra of stellar populations as a function of different chemical mixtures.
At present, several sets of stellar population models allow us to predict the sensitivity of absorption features to variations in individual elemental abundances. A first widely used framework is that adopted by the \textit{alf} code (\citealt{conroyvandokkum2012}; \citealt{conroy2018metal}), which relies on empirical stellar libraries adjusted through synthetic response functions that capture the fractional spectral changes induced by varying each element as a function of age and metallicity. A second well-established set is that of \cite{thomas2011} (hereafter TMJ), which provides predictions for Lick absorption-line indices with variable abundance ratios, based on the evolutionary population synthesis models of \cite{maraston1998evolutionary} and \cite{maraston2005}.

In addition to these widely used model families, \cite{knowles2023} provide updated response functions for a broad set of elemental abundance variations, based on the \cite{vazdekis2015} general methodology. Other model suites focus instead on specific chemical species: for instance, \cite{la2017imf} presented models tailored to exploring variations in the [Na/Fe] ratio. Finally, the fully theoretical models by \cite{coelho2014} and \cite{worthey2011} provide a self-consistent synthetic approach that covers abundance variations across the spectrum without relying on empirical stellar libraries.

Looking ahead, we expect that more accurate models will be developed in our desired rest-frame optical wavelength range thanks to the availability of new stellar libraries that include stars with variable chemical abundances (e.g. the MaStar library, \citealt{mastar}; \citealt{maraston2020stellar}; \citealt{hill2022}).

\subsection{Instrument requirements}\label{req}
As discussed above, the science case presented here requires high S/N spectra with adequate spectral resolution in the rest-frame optical range for galaxies at $z = 2-3$.
This corresponds to observations in the near-infrared,
extending up to the K band. At $z = 3$, this wavelength range includes the $\sim 5000$\ \AA\  region in the rest frame. This
region is particularly relevant, as it contains numerous absorption features sensitive to both iron and
magnesium, including the Mgb absorption and several Fe lines, making it particularly suitable for detailed
chemical abundance studies..
As mentioned in Section \ref{sample}, the typical size of such galaxies is of the order of $0.1$ arcsec, thus requiring an instrument with comparable spatial resolution.
In the following section, we will show how the unique characteristics of SHARP make it the ideal instrument to address this proposed science case.

\section{SHARP as the ideal instrument}\label{sharp}
The first requirement of our proposed science case is to obtain high S/N (i.e., S/N $ > 20$ per resolution element) on the continuum spectra of targets that are about $0.1$ arcsec wide and brighter than K$_{AB} = 24$.
A second requirement to be considered is that this S/N level should be achievable with a total exposure time not exceeding  $\sim 30$ hours. In fact, when relying on ground based telescopes, longer integrations are generally not effective, as systematic effects, particularly those related to sky subtraction residuals, tend to dominate, preventing any further improvement in S/N (e.g., \citealt{sharpandparkinson}). And given the faintness of the targets, this goal can only be achieved with the next generation 40m class telescopes. Indeed, 10m class telescopes can reach such magnitudes only in imaging - for instance, in the GOODS-South survey with VLT/ISAAC (\citealt{Retzlaff2010}) - while the practical limit for spectroscopic observations is at least about two magnitudes brighter.

\begin{figure*}%[]
  \centering
   \includegraphics[width=15cm]{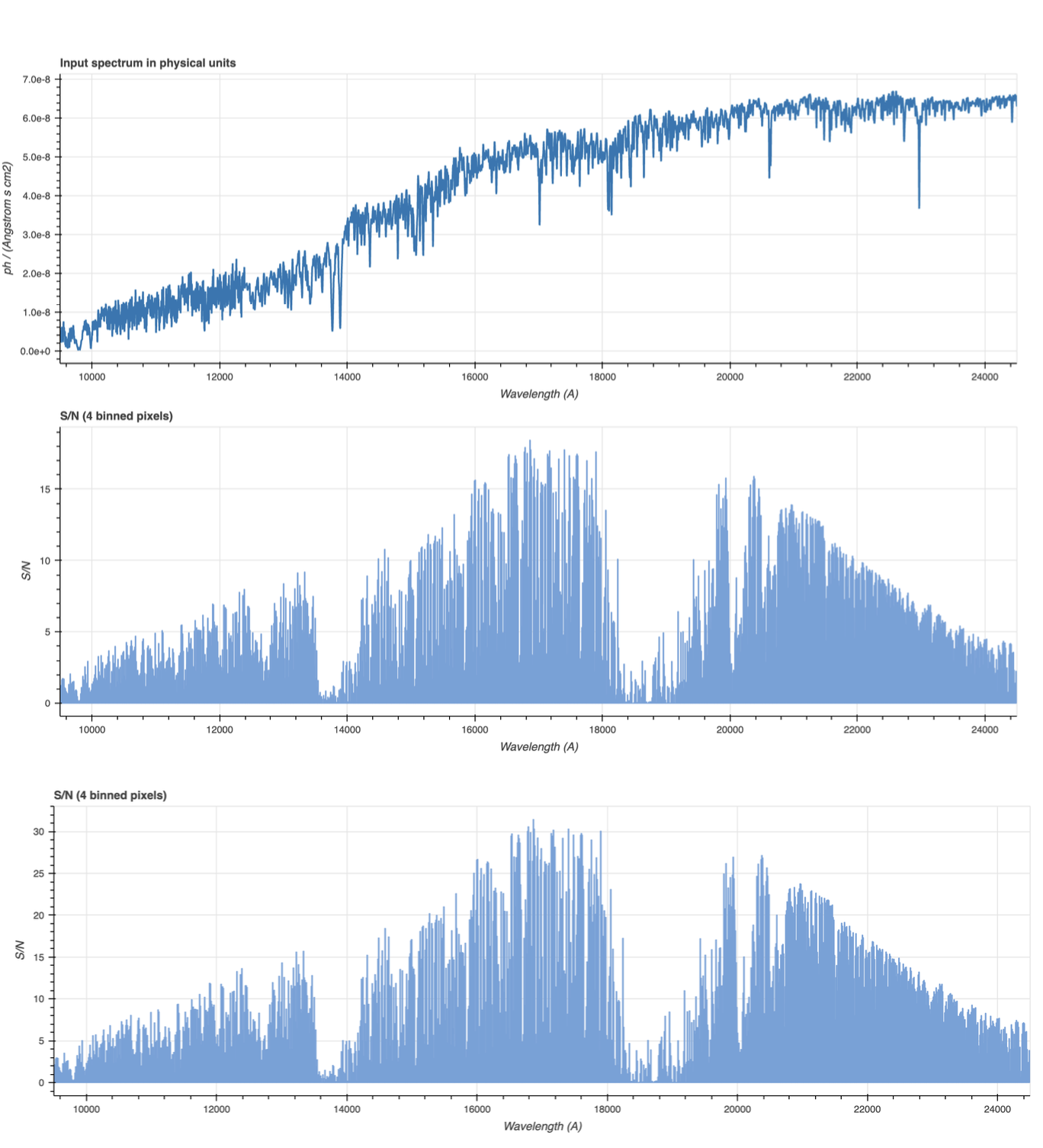} 
    \caption{Output from the ETC 0.6 of SHARP at the forthcoming ELT telescope (Franzetti et al., in prep.). \textit{Upper panel:} the input spectrum in physical units, modelled as an SSP of 2 Gyr and solar metallicity at $z=2.5$; \textit{middle panel:} S/N per resolution element obtained with 30 hours of effective esposure time using GLAO; \textit{lower panel:} S/N per resolution element obtained with 30 hours of effective esposure time using MCAO.}
    \label{snfig2}
\end{figure*}

In the following, using the ETC 0.6 $\beta$ testing version for SHARP at the forthcoming ELT telescope (Franzetti et al., in prep.), we demonstrate that a powerful adaptive optics system (e.g., Multi Conjugate Adaptive Optic - MCAO hereafter, 
https://elt.eso.org/instrument/MORFEO/, \citealt{ciliegi24morfeo}) is required to achieve both the necessary S/N and the required spatial resolution ($\sim$ 0.1 arcsec) within a total exposure time of $\sim $ 30 hours. In Figure \ref{snfig2} we present the S/N results obtained after 30 hours of effective exposure time using SHARP at the ELT with Ground Layer Adaptive Optic (middle panel, GLAO hereafter, see https://elt.eso.org/telescope/adaptiveoptics/) and MCAO (lower panel) for a galaxy simulated as a 2 Gyr, solar metallicity SSP at $z=2.5$ (upper panel). The simulated galaxy has an effective radius of $0.1$ arcsec, Sérsic index of $4$, and a total K-band magnitude of 24. 
Simulations assume a slit width of 200 mas at the nominal spectral resolution R $= 2000$. Since the absorption features we aim to measure in the galaxy continuum are broad, we consider the S/N obtained with a $\times 4$ on-chip spectral binning, performed during readout. This results in an effective pixel size of $4 \times 2.75$ \AA, while keeping the read-out noise at the level of a single (binned) pixel. 
%The target galaxies are so massive that this relatively low effective resolution has no significant impact on the resulting spectra, given that their expected stellar velocity dispersions are around 300 km s$^{-1}$, corresponding to R$_{\sigma gal} \approx$ 450. 
The number of spatial pixels used to compute the S/N has been fixed to 6 (i.e., 200 mas), matching the adopted slit width, in order to include the total flux within an area corresponding to the galaxies effective radii.

% Figure

From Figure \ref{snfig2} we see that, using a GLAO system (i.e. the standard AO mode of the ELT), the S/N reached at $\lambda \simeq 5000$\ \AA\ in the rest frame after 30 hours of exposure would be between 15 and 20 per resolution element. The lower panel of Figure \ref{snfig2} highlights the gain provided by the use of an MCAO system, such as the one delivered by MORFEO (https://elt.eso.org/instrument/MORFEO/, \citealt{ciliegi24morfeo}) and for which SHARP has been specifically designed. In this case, the achieved S/N is significantly higher ($>20-25$), allowing either more efficient observations or the possibility to target fainter magnitudes and lower-mass galaxies.
While S/N values of $15-20$ are sufficient to derive the main stellar population parameters, constraining individual elemental abundances requires substantially higher S/N. Such performance can only be reached through the MCAO capabilities of MORFEO and the NEXUS@SHARP instrument.

Summarizing, the proposed science case requires a 40-m class telescope equipped with an efficient adaptive optics system operating across the full near-IR range in order to achieve the desired goals. At the time of writing, no observing facility with these characteristics is foreseen. In particular, the ELT will host the MOSAIC multi-object spectrograph which, although it will possibly exploit the ELT GLAO system, is a fiber-fed instrument and therefore generally less efficient than slit-based systems. Furthermore, its expected wavelength coverage will extend only up to the H band, leaving the K band unexplored. This implies that key absorption features such as the Fe and Mgb indices will be inaccessible for galaxies at the higher end of our redshift range ($z \sim 3$). Moreover, without K-band coverage, even galaxies at $z \lesssim 2.7$ would lack access to the NaD index. Even though recent studies of massive quiescent galaxies at z>2 have shown that this feature is particularly susceptible to contamination by interstellar neutral gas absorption (\citealt{davies2024outflowNaD}, \citealt{taylor2026outflowNaD}, \citealt{sun2026outflowNaD}) requiring careful modelling and joint kinematic decomposition to disentangle the stellar and interstellar contributions for accurate abundance measurements, it also offers a compelling new scientific opportunity for SHARP-NEXUS. In particular, frequently detected blueshifted NaD profiles provide a direct means of characterizing massive outflows in the first generation of quiescent galaxies (\citealt{Belli2024NatureOutflow}, \citealt{deugenio2024NatAsOutflow}).

In addition, SHARP-NEXUS offers a further advantage over MOSAIC: its innovative design allows the slit to be oriented along the major axis of each target galaxy, thereby maximising the collected flux and optimising the overall S/N of the spectrum. 
%This capability, made possible by the novel architecture of the instrument (see Saracco et al., this volume), is not achievable with MOSAIC’s fiber based approach. 
Massive quiescent galaxies at z$\sim$2 are notoriously compact, exhibiting a typical effective radius of $R_{e} \sim$ 0.1 arcsec but featuring high ellipticities (q $\approx$ 0.6) and extended wings typical of high Sérsic indices (\citealt{ito2024}). In observations that are already at the feasibility limit to achieve a sufficient S/N, this unique capability of SHARP-NEXUS provides up to a 30\% increase in collected stellar flux compared to MOSAIC’s fixed-aperture fibers (0.6" - 0.8"), which inevitably truncate the light profile along the major axis. By aligning the slit with the galaxy's morphology, SHARP-NEXUS maximizes the source signal without unnecessarily increasing the sky background contribution.

\section{Observing strategy}\label{obsstra}
In Section \ \ref{sample}, we described the characteristics of the typical target of our proposed case. Here we describe the observing strategy we want to propose. 

The target selection will start from parent samples with multi-band photometric catalogues providing reliable redshift estimates, preferably spectroscopic, or photometric when spectroscopy is unavailable. A K-band magnitude cut ($K_{AB}$ < 24) will then be applied. The quiescent nature of the candidate massive galaxies will subsequently be assessed through a colour-based pre-selection using their rest-frame U-V versus V-J colours (\citealt{williams2009}) or the $(ugi)_{s}$ selection of \citealt{Antwi-Danso2023}.
The volume density of all galaxies with $log(M_\star/M_\odot) > 10.5$ (i.e., $K$-band (AB) $< 24$) between $z=2$ and $z=3$  is $\phi \sim 5 \times 10^{-4}$ Mpc$^{-3}$ (\citealt{muzzin2013}; \citealt{behroozi2013}). Quiescent systems represent roughly $15\%$ of this population (see Fig. $6$ in \citealt{muzzin2013}). This implies a quiescent number density of $\phi_Q \sim 7.5 \times 10^{-5}$ Mpc$^{-3}$, corresponding to about 1000 possible targets over $1$ deg$^2$ (i.e. 0.3 objects per arcmin$^{-2}$). In this program we propose to sample approximately $25\%$ of the total massive quiescent population across $1$ deg$^2$ field, corresponding to a complete, mass-limited sample of $\sim 200$ galaxies. 

Robustly selected catalogues of high-redshift massive quiescent galaxies are already available in the literature, such as the sample of 174 massive quiescent galaxies in the COSMOS field presented by \cite{deugenio2021} and \cite{valentino2023}. JWST legacy fields such as JADES (\citealt{Eisenstein2026jades}) also provide deep, multi-band photometry suitable for reliable SED fitting and the identification of further candidates (e.g. Robertson et al. 2026). These existing samples will be further expanded by forthcoming wide-area surveys such as Euclid (\citealt{Euclid_paper1_Mellier}) and the Vera C. Rubin Observatory (LSST), providing an even larger and more homogeneous pool of potential targets by the time of the SHARP observations.

Given the low surface density of these objects, we expect to occupy only one of the 30 slits available per NEXUS pointing over $1.2 \times 1.2$ arcmin$^{2}$ field. The sparse distribution of our target galaxies implies that, on average, only a single target per field will fall within the accessible area. 
As a consequence, we anticipate sharing the focal plane with other surveys interested in higher number density targets, for example programmes targeting star-forming galaxies at similar redshifts. Such coordinated strategies would ensure an efficient use of the NEXUS multiplexing capabilities. This approach has already been successfully implemented in surveys based on much larger target samples. A clear example is provided by the 4MOST instrument on ESO’s VISTA telescope (\citealt{dejong2019}), which, thanks to its 2400 fibre multiplexing over a 3 deg$^2$ field of view, enables focal plane sharing among up to 25 survey programmes (\citealt{tempel2020}).

To the field galaxy sample, we plan to add an additional set of approximately 200 quiescent galaxies located in denser environments. Even without accounting for the structures that will likely be uncovered at $z>2$ by ongoing photometric surveys (e.g. Euclid, see \cite{Euclid_paper1_Mellier} and \cite{euclid_overdensities_2025}), several galaxy overdensities have already been discovered, whose stellar populations clearly indicate that they are clusters at $z \sim 2$ or higher.
 An example is JKCS-041, a galaxy cluster at $z=1.8$ (\citealt{Andreon2009JKCS041}), where the already spectroscopically confirmed quiescent members are sufficiently numerous that a single SHARP-NEXUS pointing would include at least ten cluster galaxies, plus a few foreground systems. Another example is CL1 in the XMM-LSS field, identified through Hyper Suprime-Cam photometry at 
z$=2.09$, which contains about ten quiescent galaxies within the SHARP-NEXUS field of view (\citealt{Kiyota}). Similarly, XLSSC122, observed with HST/WFC3 (F105, F140, and G141), lies at z$=1.98$
and hosts approximately fifteen red-sequence galaxies within the SHARP-NEXUS field of view (\citealt{willis}). Even more striking is the recent discovery of a protocluster at 
z$=3.37$ featuring a remarkably high fraction of quiescent galaxies (\citealt{McC}). 
As a further check on the selection of quiescent galaxies in overdense environments, the UVJ colours of the target galaxies (already available for all systems except the z=3.37 protocluster studied by \citealt{McC}), will be will be examined to identify the quiescent population. This will ensure that the resulting sample is selected consistently with, and is therefore directly comparable to, the field sample.

It is worthy to note that investigating the stellar population properties of high redshift massive quiescent galaxies in dense cluster environments, and comparing them with their field counterparts, is crucial for constraining the role of environmental quenching mechanisms. At intermediate redshifts, galaxies residing in dense cosmic environments and groups have been found to exhibit older stellar populations and shorter star-formation timescales (e.g., \citealt{ditrani2025}). Observations at higher redshifts therefore provide a unique opportunity to test whether these environmental trends were already in place at earlier cosmic epochs.

For this part of the sample, we expect to need approximately 20 pointings to obtain a dense environment subsample of about 200 quiescent galaxies.
Therefore, the proposed survey will require approximately 220 pointings, each with 30 hours of exposure. The total observing time corresponds to about 12\% of the available ELT time over five years, assuming a multi-survey programme with shared use of the focal plane.

It is worth emphasising that the proposed baseline of 220 pointings represents the ideal scenario for carrying out a statistically robust study of environmental quenching across different cosmic epochs. However, if focal-plane sharing proves impractical or the available time budget is limited, a more focused and realistic alternative would be to frame the project as a typical ESO Large Programme with a total allocation time of 300 hours (10 pointings) over 2-3 years, adopting a 2-cluster and 8-field configuration. The two cluster pointings would provide a high yield sample of dense environments, while the eight field pointings would target galaxies with existing JWST observations. This configuration would provide a direct empirical benchmark for the stellar population parameters derived from SHARP-NEXUS and allow us to assess the additional information provided by high quality ground-based spectroscopy compared to current space-based observations. Although this reduced programme would not provide the same statistical precision as the full baseline, it would retain the core scientific objectives while offering a realistic and competitive path towards implementation.

\section{Summary and conclusion}\label{summary}

Massive quiescent galaxies provide a unique window into galaxy formation and evolution. %Massive quiescent galaxies at cosmic noon represent a cornerstone population for our understanding of galaxy formation and evolution. 
Their existence at $2 < z < 3$, already assembled, metal-rich, and passive, poses a long-standing challenge to hierarchical formation models in a cold dark matter Universe. At the same time, their early assembly and subsequent slow evolution make them powerful and precise cosmological probes (e.g., BOSS, \citealt{eisenstein2001}; \citealt{eisenstein2005}). Directly probing their stellar population properties and detailed element abundances close to the epoch of quenching is therefore essential to disentangle formation pathways from subsequent evolutionary processes such as mergers and progenitor bias.

The combination of the ELT aperture, the MORFEO MCAO system, and the SHARP-NEXUS instrument uniquely enables this science case. Its slit-based design, high throughput, large near-IR wavelength coverage, and capability to align the slit along the galaxy major axis provide the sensitivity and angular resolution needed to access key absorption features to measure individual element abundances in galaxies observed during the main epoch of quenching. At present, no other foreseen facility offers a comparable combination of performance and spectral coverage for this science case.

Although massive quiescent galaxies are intrinsically rare at these redshifts, we showed that an efficient observational strategy is feasible by exploiting SHARP-NEXUS multiplexing through coordinated surveys and by targeting known galaxy overdensities at $z > 2$. Such synergistic approaches, already proven effective by large multiplex facilities (such as 4MOST), allow the construction of statistically meaningful, mass-complete samples spanning both field and dense environments within a realistic time allocation.

SHARP-NEXUS can open a fundamentally new observational window on the chemical enrichment and star formation histories of the earliest massive quiescent galaxies. By directly measuring their stellar elemental abundances at the epoch of formation, this programme would provide critical constraints on galaxy formation models that cannot be obtained by any other planned facility, firmly establishing the role of ELT-class spectroscopy in unveiling the origin of today’s massive early-type galaxies.

\printcredits

\section*{Acknowledgements}
We warmly thank both anonymous referees for their insightful comments and suggestions, which have greatly improved the readability of the paper and the completeness of its content. ML, FD, SZ, AGal and FLB acknowledge financial support from INAF Large Grant 2022, FFO 1.05.01.86.16, and the Italian Ministry grant Premiale MITIC 2017.
FLB acknowledges support from INAF 2023 RSN1 Minigrant 1.05.23.04.01 and  from the Fundación Occident and the Instituto de Astrofísica de Canarias under the Visiting Researcher Programme 2022–2025 agreed between both institutions. PS and the SHARP team acknowledge support from the INAF Techno-Grant 2022, 1.05.12.02.01, and INAF Large-Grant 2024, 1.05.24.01.01. ML acknowledges support from INAF 2022 RSN1 Minigrant (1.05.12.04.01) and INAF 2024 RSN1 Minigrant (1.05.24.07.01).
.


% Use the supplied, complete bibliography directly for arXiv.
% No BibTeX run is required. For regeneration, see the separate build notes.



\end{document}